\documentclass[a4paper]{jpconf}
\usepackage{epsfig}      
\usepackage[section]{placeins}

\begin{document}
\title{Study of the PYTHIA $\Delta\eta\Delta\phi$ correlation as a function of multiplicity and transverse sphericity in proton-proton collisions at 7 TeV.}

\author{E. P\'erez Lezama, G. Pai\'c, E. Cuautle Flores, A. Ortiz Velasquez}

\address{ Instituto de Ciencias Nucleares, Universidad Nacional Aut\'onoma de M\'exico Apartado Postal 70-543, 04510 M\'exico, D. F., M\'exico}

\ead{edgar.perez.lezama@cern.ch} 

\begin{abstract}
The study and understanding of the details of proton-proton collisions is important since they are the benchmark for the comparison with heavy ion results. In the present work we have studied the $\Delta\eta\Delta\phi$ correlations as a function of the event multiplicity and transverse sphericity in pp collisions at $\sqrt{s}$ = 7 TeV using PYTHIA 8.180. The results show large variation in the shape of the correlation as a function of the sphericity and multiplicity indicating the importance of slicing the data in bins of both observables to better understand the complexity of the interactions. Motivated by the radial flow patterns originated by color reconnection (CR), results with and without CR together with dihadron correlations are discussed.
\end{abstract}

\section{Introduction}
The pp collisions have a complicated structure which involves complex final states, with large multiplicities of hadrons, all originated from different processes \cite{ALICEmult}, this complexity of the interactions requires MC event generators to simulate them. PYTHIA\cite{pythia} is an event generator which  comprises a coherent set of physics models for the evolution from a few-body hard process to a complex multihadronic final state. Partonic processes can be classified as being either soft or hard.
 
Different techniques of analyses are used to investigate the properties of the hadronic interactions, among them the dihadron azimuthal correlations. Using this observable, an interesting ridge-shaped correlation has been discovered in collisions of heavy nuclei. Some models attribute the ridge to jet-medium interactions, while others attribute it to the medium itself \cite{Ridge1}-\cite{Ridge8}. Recent results from CMS have revealed a striking ridge structure in very high multiplicity proton-proton (pp) collisions at a center-of-mass energy of 7 TeV. This novel structure resembles similar features observed in relativistic heavy-ion experiments \cite{Ridge1}. Several mechanisms like Color-Glass Condensate based on initial state nonlinear gluon interactions\cite{CGC} and elliptic flow\cite{ellipticflow} have been proposed to explain the phenomenon.

Although the effect observed by CMS, cannot be reproduced\cite{Ridge8} by PYTHIA 8.180, in this work we investigate the possible presence of the ridge, studying the dihadron correlation as a function of the event shape variable, since it allows to control the amount of MPI and hardness of the partonic interactions\cite{NewPaper}. According with Ref.\cite{CR5} the color reconnection mechanism (CR) produces radial flow effects in events with a large number of multi-partonic-interactions and the same effect  might produce elliptic flow.

At hadron colliders the event shapes are calculated in the transverse plane in order to avoid the bias from the boost along the beam axis. The transverse sphericity $(S_{\rm T})$ is defined in terms of the eigenvalues $\lambda_{1}>\lambda_{2}$ of the transverse momentum matrix:

\begin{equation}
S_{xy} = \frac{1}{\sum_{i}p_{\rm T_{i}}}\sum_{i}\frac{1}{p_{\rm T_{i}}} 
\left( \begin{array}{cc}
p_{x_{i}}^{2} & p_{x_{i}}p_{y_{i}} \\
p_{y_{i}}p_{x_{i}} & p_{y_{i}}^{2}
\end{array} \right)
\end{equation}

\noindent in the following way

\begin{equation}
S_{\rm T} = \frac{2\lambda_{2}}{\lambda_{2}+\lambda_{1}}.
\end{equation}

\noindent The limit values of this variable are related to specific configuration in the transverse plane

\[ S_{\rm T} = \left\{ \begin{array}{ll}
0 &  \mbox{``pencil like" limit} \\
1 &  \mbox{``isotropic" limit}
\end{array} \right. \]

\section{Analysis Details}
Using PYTHIA 8.180 we generated 850 million events with CR and the same without CR. In order to have a meaningful sphericity, the observable is calculated for events with more than two primary charged particles. They are defined as all those produced in the collisions including weak and electromagnetic decays products, but excluding the products from weak decays of strange particles. Only particles with $p_{\rm T}>0.5$ GeV/c and $|\eta|<1$ are considered in the results. The leading particle is selected in the pseudo-rapidity range $\mid\eta\mid<1.0$, the same as sphericity, and having the $\eta$ range open for the associated particles. In this work the trigger and associated particles are not selected in a certain momentum range, as done in the conventional dihadron correlation analyses. The leading particle is the one with the highest transverse momentum in each event, and the associated particles are all the others that fulfil the $p_{\rm T}$ and $\eta$ cuts previously described. Thus, the correlation is built by taking the differences in $\eta$ and $\phi$ between the leading and the associated particles:
\begin{equation}
\begin{array}{ccc}
\Delta\eta & = & \eta^{assoc} - \eta^{lea} \\
\Delta\phi & = & \phi^{assoc} - \phi^{lea} \\

\end{array}
\end{equation}

\noindent Where $\eta^{lea}$ and $\phi^{lea}$ ($\eta^{assoc}$, $\phi^{assoc}$) are the pseudo-rapidity and the azimuthal angle of the leading particle (associated), respectively. 

\begin{figure} [t]
\begin{center}
\epsfig{file= 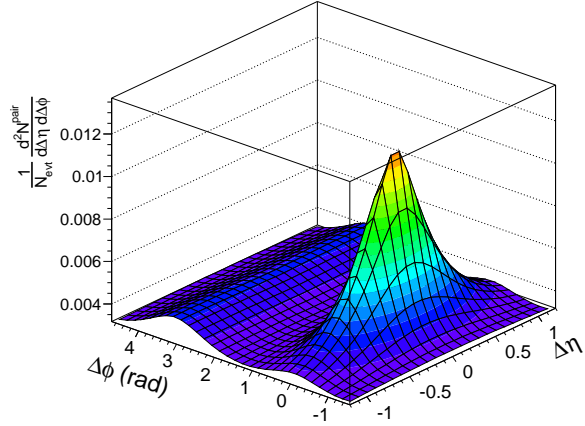,bb=0 0 567 474 ,scale=0.4}
\caption{$\Delta\eta-\Delta\phi$ correlation for a leading and associated particle selected in the region $|\eta|<1.0$, where multiplicity and sphericity cuts are not applied.}
\label{ima::DihCorr}
\end{center}
\end{figure}

 A plot of the dihadron correlations for minimum bias events is shown in figure \ref{ima::DihCorr}.   Two prominent structures are visible, one at $(\Delta\phi,\Delta\eta)\approx0$, known as ``near-side peak", and correspond to correlations from single jets. The second structure the``away-side peak" is present at $\Delta\phi\approx\pi$ and corresponds to the recoil jet.

\section{Results}
The results are presented using two sphericity ranges, $S_{T}<0.1$ and $S_{T}>0.9$, corresponding to jetty-like and isotropic events, respectively.   The correlations are computed as a function of the event multiplicity.   The multiplicity is expressed in terms of $z=(dN_{ch}/d\eta)/\langle dN_{ch}/d\eta \rangle$, i.e. the ratio of the multiplicity and the mean multiplicity. The multiplicity ranges presented are: $z<1$, $2<z<3$ and $4<z$. The correlations are normalized to the number of leading particles(same as number of events) in each sphericity and multiplicity range being analyzed.

\begin{figure} [h]
\begin{center}
\epsfig{file= 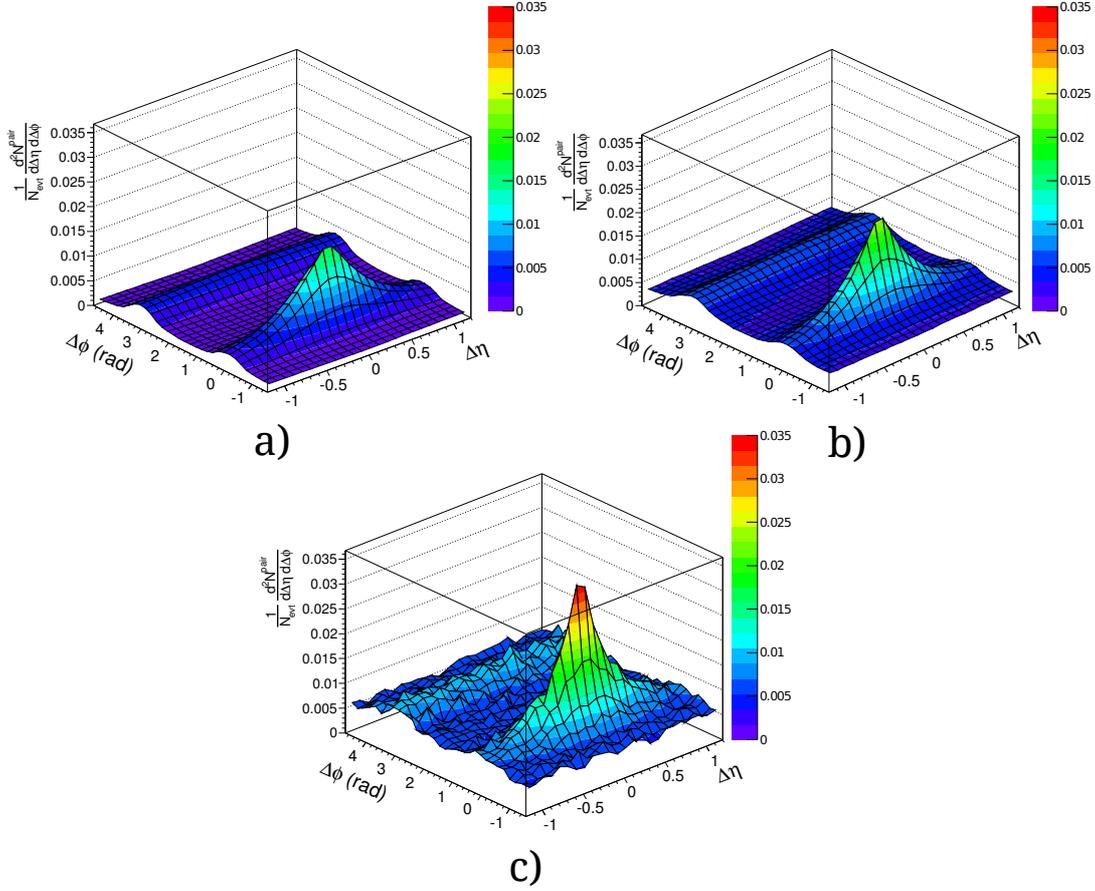,bb=0 -1 1190 977 ,scale=0.34}
\caption{Dihadron correlations for low sphericity events $(S_{\rm T}<0.1)$ in three multiplicity ranges, CR is activated.(a) for $z<1$; (b) for $2<z<3$ and (c) for $4<z$ .  }
\label{ima:LowSt}
\end{center}
\end{figure}

Figure \ref{ima:LowSt} shows the dihadron correlation for low sphericity events ($S_{\rm T}<0.1$), where no ridge proper structure has been observed in the ranges studied, although an extended structure of the near side peak is observed, that will be the subject of further studies.

\FloatBarrier

\begin{figure} [h]
\begin{center}
\epsfig{file= 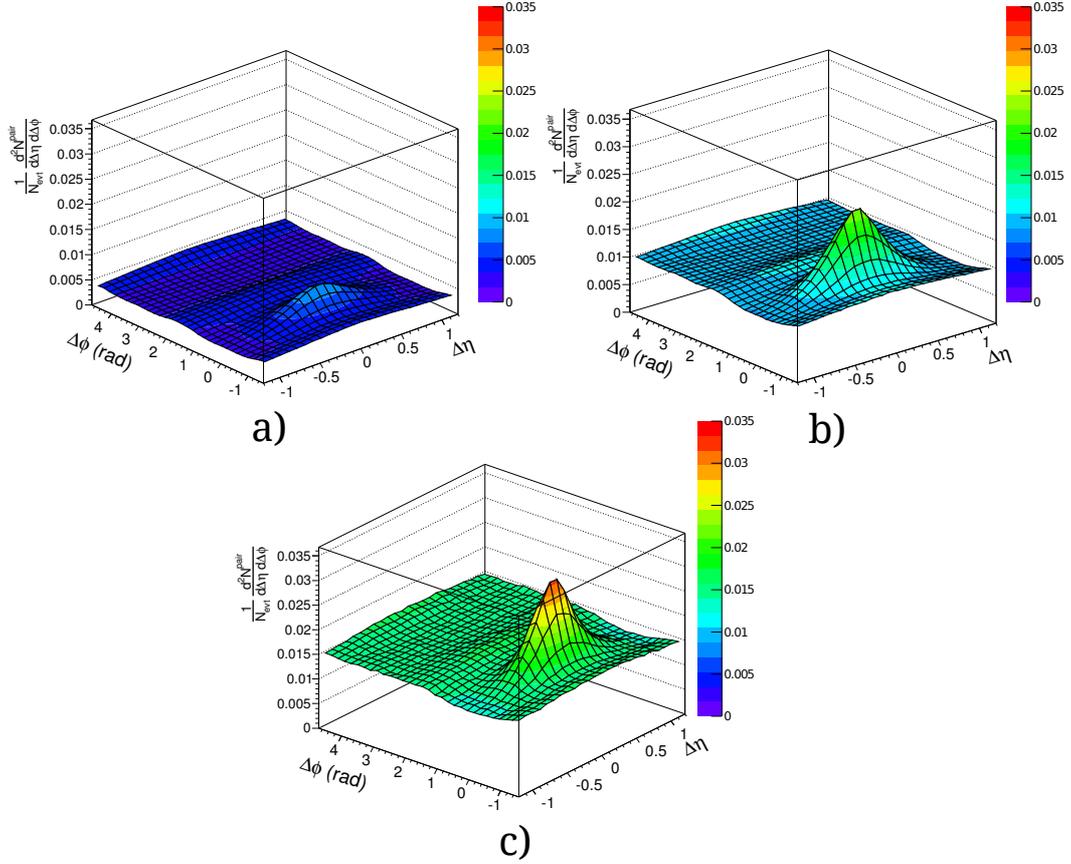,bb=0 -1 1195 977 ,scale=0.33}
\caption{Dihadron correlations for high sphericity events $(S_{\rm T}>0.9)$ in three multiplicity ranges, CR is activated, (a) for $z<1$; (b) for $2<z<3$ and (c) for $4<z$.  }
\label{ima:HighSt}
\end{center}
\end{figure}

Figure \ref{ima:HighSt} shows the dihadron correlations for $S_{\rm T}>0.9$ in three multiplicity ranges, where an interesting three hump structure appears, even though in PYTHIA the $2\rightarrow3$ processes are only indirectly included, through final state radiation, since PYTHIA is a leading order (LO)event generator. Similar structures are present when color reconnection mechanism is switched off, so the shape is not due to a flow effect. This has been already reported in \cite{Paper}.

\begin{figure} [h]
\begin{center}
\epsfig{file= 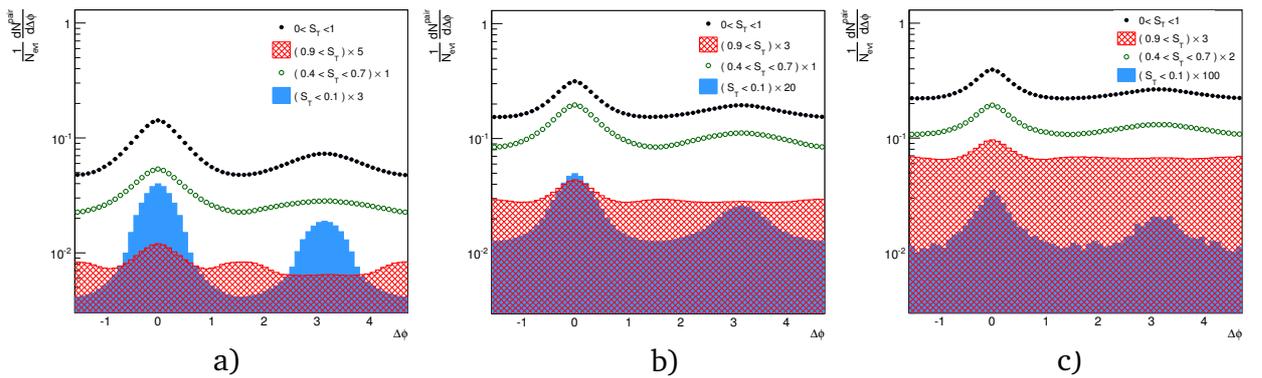,bb=0 -1 1613 507 ,scale=0.3}
\caption{Projection in the $\Delta\phi$ axis selecting three $S_{\rm T}$ regions and three multiplicity ranges: (a) for $z<1$; (b) for $2<z<3$ and (c) for $4<z$. Black points represent the events with no sphericity cut.}
\label{ima:ProjCR}
\end{center}
\end{figure}

The projections on the $\Delta\phi$ axis of the dihadron correlations, normalized to the number of events in the multiplicity range, show the underlying structures for three different ranges in the transverse sphericity. In the jetty events $(S_{\rm T}<0.1)$ we observe the expected near and away-side peaks, for high sphericity $(0.9<S_{\rm T})$   we observe the  double hump . For the intermediate sphericity $(0.4<S_{T}<0.7)$ we observe a broad away side with some structure. Figure \ref{ima:ProjCR} shows these projections together with  the black curve representing the projection of  all events in a given multiplicity range. We can see that the total correlation shape is a sum of different structures, that are identified by the use of the sphericity variable.

\FloatBarrier

\section{Conclusions}
In this work  we have introduced transverse  sphericity to analyze dihadron correlations. By using this event shape variable, we were able to analyze different types of events like those dominated by jetty like and/or isotropic structures.  In high sphericity events we observe clearly 3 track structures forming a ``T", especially at low multiplicities (figure \ref{ima:HighSt}). The ``T" structure manifest itself by the leading particle peak at $\Delta\phi\approx0$ and the two humps at $\Delta\phi\approx\pi/2$ and $\Delta\phi\approx 3\pi/2$. We have shown that by sectioning p-p events in ranges of sphericity one uncovers details of the so called underlying event\cite{UE}, usually treated as an amorphous structure.

\ack
Partial support for this work has been received by CONACyT under the grant numbers 103735 and 101597; and PAPIIT-UNAM under the projects: IN105113 and IN108414.


\section*{References}
\begin{thebibliography}{9}

\bibitem{ALICEmult} ALICE collaboration, K Aamodt et al., 2010  \textit{Eur. Phys. } J \textbf{C68} 345

\bibitem{pythia} T Sjostrand, P Eden, C Friberg, L Lonnblad, G Miu, S Mrenna and E Norrbin, 2001 \textit{Computer Physics Commun} \textbf{135} 238.

\bibitem{Ridge1} STAR collaboration, B I  Abelev et al., 2009  \textit{Phys. Rev.} \textbf{C80} 064912 .

\bibitem{Ridge8} D Velicanu, 2011  \textit{J. Phys. } G \textbf{38} p. 124051


\bibitem{CGC} K Dusling, R Venugopalan, 2013  \textit{Phys. Rev.} D \textbf{87} 094034.

\bibitem{ellipticflow} K Werner, I Karpenko and T Pierog 2011  \textit{Phys. Rev. Lett.} \textbf{106} 122004




\bibitem{NewPaper} E Cuautle et al, Disentangling the soft and hard components of the pp collisions using the sphero(i)city approach 2014 arXiv:1404.2372

\bibitem{CR5} T Sj\"ostrand, S Mrenna and P Skands 2006 \textit{JHEP 0605} \textbf{026}


\bibitem{Paper} A Ayala et al 2009  \textit{Eur. Phys.} J \textbf{C62} arXiv:0902.0074

\bibitem{UE} R Field, 2012   \textit{Acta Phys. Polon.} B \textbf{42} 2631 arXiv:1202.0901

\bibitem{CMSpythia} S Chatrchyan, CMS Collaboration, Long-range and short-range dihadron angular correlations in central PbPb collisions at a nucleon-nucleon center of mass energy of 2.76 TeV  2011 \textit{JHEP} \textbf{1107} 076 arXiv:1105.2438


\end{thebibliography}
\end{document}